\documentclass[11pt]{article}
\pdfoutput=1

\usepackage{amsmath}
\usepackage{amssymb}
\usepackage[show]{chato-notes}
\usepackage{graphicx}

\usepackage{cite}
\usepackage{placeins}

\usepackage[labelfont=bf,labelsep=period,justification=raggedright]{caption}
\usepackage{caption}
\usepackage{ulem}

\makeatletter
\renewcommand{\@biblabel}[1]{\quad#1.}
\makeatother

\date{}

\begin{document}

\begin{flushleft}
  {\Large \textbf{World impact of kernel European Union 9 countries \\
     from  Google matrix analysis of the world trade network} }
\\ \medskip
Justin Loye$^{1,2}$,
Leonardo Ermann$^{3,4}$,
Dima L.\ Shepelyansky$^{1,*}$
\\ \medskip
{1} \newblock { Laboratoire de Physique Th\'eorique du CNRS, IRSAMC,
Universit\'e de Toulouse, CNRS, UPS, 31062 Toulouse, France}
\\
{2} \newblock { Institut de Recherche en Informatique de Toulouse, 
		Universit\'e de Toulouse, UPS, 31062 Toulouse, France}
\\
{3} \newblock { Departamento de F\'{\i}sica Te\'orica, GIyA,
 Comisi\'on Nacional de Energ\'{\i}a At\'omica.
 Av.~del Libertador 8250, 1429 Buenos Aires, Argentina}
\\
{4} \newblock {  Consejo Nacional de Investigaciones
Cient\'ificas y T\'ecnicas (CONICET), Buenos Aires, Argentina}
\\
\medskip
$\ast$ E-mail: dima@irsamc.ups-tlse.fr
\end{flushleft}

\section*{Abstract}
We use the United Nations COMTRADE database for analysis of the multiproduct world trade
network. With this data, considered for years 2012-2018,
we determined the world trade impact of the Kernel of EU 9 countries (KEU9), 
being Austria, Belgium, France, Germany,
Italy, Luxembourg, Netherlands, Portugal, Spain,
considered as one united country. We apply the advanced Google matrix analysis
for investigation of the influence of KEU9
and show that KEU9 takes the top trade network rank positions
thus becoming the main player of the world trade being ahead of USA and China.
Our network analysis provides additional mathematical
grounds in favor of the recent proposal \cite{stetienne} 
of KEU9 super-union which is  based
only on  historical, political and economy basis.

\noindent
{\bf Dated: October 21, 2020}

\newpage$\phantom{.}$

\section*{Introduction}

The economy of European Union (EU) is considered
as the second world largest economy after United States (US) \cite{wikieconomy}
even if there are also other opinions placing China (CN) 
on the first position \cite{china1st}. At present EU includes 
27 member states and about 447 million population \cite{wikieu}.
While the global EU economy and population are really huge
the political action of member states \cite{wikieupolit} is not always coherent 
pushing in some cases in different directions.
Due to this reason there is a proposal, pushed forward by Christian Saint-Etienne, 
to consider the Kernel EU 9 (KEU9) states (or countries),
which are tightly linked by historical, political and economic relations,
as a strongly united kernel group of EU
that would allow to perform coherent actions 
of these KEU9 states \cite{stetienne} (follow also discussion of this proposal
at \cite{zimour}).
These 9 kernel states include
Austria (AT), Belgium (BE), France (FR), Germany (DE),
Italy (IT), Luxembourg (LU), Netherlands (NL),
Portugal (PT), Spain (ES) \cite{stetienne} with the total population of 
about 305 millions \cite{wikieu}.

A variety of arguments in favor of possible coherent political and economic actions of KEU9 group
is presented and analyzed in \cite{stetienne}. However, this analysis is
not based on detailed mathematical grounds pushing forward
arguments of  historical, political  and economic heuristic reasons.
Here, we put forward the mathematical foundations for this KEU9
proposal presenting the mathematical and statistical
analysis of the world trade database of UN COMTRADE \cite{comtrade}.
This database presents an exceptional variety of data on 
trade exchange between all world UN registered countries
on a scale of more than 50 years with 
more than $10^4$ of trade commodities (products). 
The transactions are expressed in their dollar (USD) 
values of a given year. 
The World Trade Organization (WTO) Statistical Review 2018 \cite{wto2018}
demonstrates the vital importance of 
the international trade between  countries 
for their development and progress.
Also the whole world economy is deeply influenced 
by the world trade \cite{krugman2011}. 
Hence, this database is well appropriate for 
verification of how strong and important is
the world influence of KEU9 group on the world
economy. Thus here we use the UN COMTRADE database \cite{comtrade}
for mathematical and statistical analysis of
heuristic arguments presented in favor of KEU9 in \cite{stetienne}.

The trade transfer between countries represents the 
multiproduct World Trade Network (WTN). 
The modern methods of Google matrix approach
\cite{brin,meyer,rmp2015} are well suited for the analysis of 
transactions on the WTN. The detailed description
of Google matrix  applications to
WTN are described in \cite{wtn1,wtn2,wtn3,wtn4}.
Here we apply these methods considering KEU9 countries as one 
country thus excluding trade transfers between them and
keeping only ingoing and outgoing trade flows
to this group from other countries.

We point that various research groups investigated the statistical properties of WTN
(see e.g. \cite{serrano07,fagiolo09,he10,fagiolo10,barigozzi10,debenedictis11,deguchi14}).
However, as discussed in \cite{wtn1,wtn2,wtn3} the Google matrix approach
has significant advantages for analysis of weighted directed trade networks
since it takes into account multiple iterative transactions
and thus provides a new and more detailed analysis of trade influence propagation
compared to the usual approach based on export and import flows.

\section*{Materials and Methods}

\subsection*{Google matrix construction of WTN}
We consider the trade exchange between 
$N_c=186$ (185 countries + KEU9) world countries and $N_p=10$ products given by
1 digit from the the Standard International Trade Classification (SITC)
Rev. 1, and for years 2012, 2014, 2016, 2018
taken from UN COMTRADE \cite{comtrade}. These 10 products contain
all smaller subdivided specific products
which number goes up to $\sim 10^4$.
The list of  these 10 products is given in Table~\ref{tab1}.
The list of world countries is available at \cite{wtn1,wtn2}.
 Following the approach developed in \cite{wtn1,wtn2}
we obtain $N_p$ money
matrices $M^p_{c,c^\prime}$ which give 
product $p$ transfer (in USD) from country $c'$
to country $c$. The Google matrices $G$ for the direct trade flow
and $G^*$ for the inverted trade flow have the size of
$N=N_c N_p=1860$ nodes. They are constructed by 
normalization of all column of outgoing weighted links
to unity. There is also the part with a damping factor 
$\alpha=0.5$ describing random trade-surfer jumps to all nodes
with a certain personalized vector taking into account 
the weight of each product in the global trade volume.
The construction procedure of $G$ and $G^*$ is described in detail in 
\cite{wtn2,wtn3}. The general properties and various examples
of Google matrices of various networks are
given in \cite{brin,meyer,rmp2015}.

The stationary probability distribution of Markov transitions
described by the Google matrix $G$ is given by the PageRank vector
$P$ with maximal eigenvalue $\lambda=1$:
$GP=\lambda P =P$ \cite{brin,meyer}. For the inverted flow described by $G^*$ matrix
we have similarly the CheiRank vector $P^*$, 
being the eigenvector of $G^* P^* = P^*$. The importance and detailed statistical analysis of
the CheiRank vector were demonstrated in
\cite{linux}  (see also \cite{wtn1,wtn2,wikizzs}).
We define PageRank $K$ and CheiRank $K^*$ indexes
by monotonic ordering of probabilities of PageRank vector $P$ and 
of CheiRank vector $P^*$ as
$P(K)\ge P(K+1)$ and $P^*(K^*)\ge P^*(K^*+1)$ with $K,K^*=1,\ldots,N$.
By taking a sum over all products $p$ we obtain the PageRank and CheiRank 
probabilities of a given country as $P_c =\sum_p P_{cp}$ and 
${P^*}_c =\sum_p {P^*}_{cp}$ (and in a similar way
product probabilities $P_p, {P^*}_p$) \cite{wtn2,wtn3}).
From these probabilities we obtain the related indexes $K_c, {K^*}_c$.
In a similar way we define 
from import and export trade volume the
 probabilities
$\hat{P}_p$, $\hat{P}^*_p$, $\hat{P}_c$, $\hat{P}^*_c$,
$\hat{P}_{pc}$, $\hat{P}^*_{pc}$ and 
corresponding indexes
$\hat{K}_p$, $\hat{K}^*_p$, $\hat{K}_c$, $\hat{K}^*_c$, $\hat{K}$, $\hat{K}^*$
(the import and export probabilities
are normalized to  unity
via the total import and export volumes, see details in
\cite{wtn2,wtn3}). We note that qualitatively
PageRank probability is proportional to the volume of ingoing
trade flow and CheiRank respectively to outgoing flow.
Thus, approximately we can say that the
high import gives a high  PageRank $P$ 
and a high export a high CheiRank $P^*$ probabilities.

\subsection*{Reduced Google matrix}
We also use the REGOMAX algorithm described in detail in \cite{greduced,politwiki}. 
This algorithm allows  to compute efficiently a 
{\it reduced Google matrix} $G_R$ of size $N_r \times N_r$ 
that accounts all transitions of direct and
indirect pathways happening in the full Google matrix $G$ 
between $N_r$ nodes of interest.
For the selected $N_r$  nodes their PageRank probabilities 
are the same as for the global
network with $N$ nodes (up to a constant multiplicative factor
which takes into account that the sum
of PageRank probabilities over $N_r$  nodes is unity). The matrix $G_R$  can be presented as
as a sum  of three matrix components that clearly distinguish direct and indirect interactions:
$G_\mathrm{R} = G_{rr} + G_{\mathrm{pr}} + G_{\mathrm{qr}}$  \cite{politwiki}. 
Thus $G_{rr}$ is produced by  the direct links between selected $N_r$ nodes in 
the global network of $N \gg N_r$ nodes.
The component $G_{pr}$  is rather close to the matrix in which each column is
given by the PageRank vector $P_r$ (up to a constant multiplier). 
Due to that $G_{pr}$ does not give much information about direct
and indirect links between selected $N_r$ nodes. 
The most  interesting and  nontrivial  role is played
by the component $G_{qr}$, 
which accumulates the contribution of all indirect links between selected 
$N_r$ nodes appearing due to
multiple pathways via the global network of $N$ nodes. 
The exact formulas for these three components of $G_R$ are given in \cite{greduced,politwiki}.

\subsection*{Sensitivity of trade balance}

Following \cite{wtn2,wtn3},
we use the trade balance of a given country
with PageRank and CheiRank probabilities defined 
as $B_c = (P^*_c - P_c)/(P^*_c + P_c)$.
In a similar way we have from ImportRank and ExportRank probabilities
as $\hat{B}_c=  ({\hat{P}^*}_c - \hat{P}_c)/({\hat{P}^*}_c + \hat{P}_c)$.
The sensitivity of 
trade balance $B_c$ to the price of energy  or machinery can be obtained 
from the change of corresponding money volume
flow related to SITC Rev.1 code $p=3$ (mineral fuels) or $p=7$ (machinery)
by multiplying it by $(1+\delta)$, then
computing all rank probabilities and 
the derivative $dB_c/d\delta$.

The efficiency of the above Google matrix methods has been
demonstrated not only for the WTN but also for variety of other
directed networks including 
Wikipedia networks \cite{wikizzs,wikicountires,wrwu2017,wikipharma}
and biological networks of 
protein-protein interactions \cite{zinprotein1,zinprotein2}. 

\section*{Results}

\subsection*{CheiRank and PageRank of countries}

\begin{figure*}[!ht]
\begin{center}
\includegraphics[width=0.92\columnwidth,angle=0]{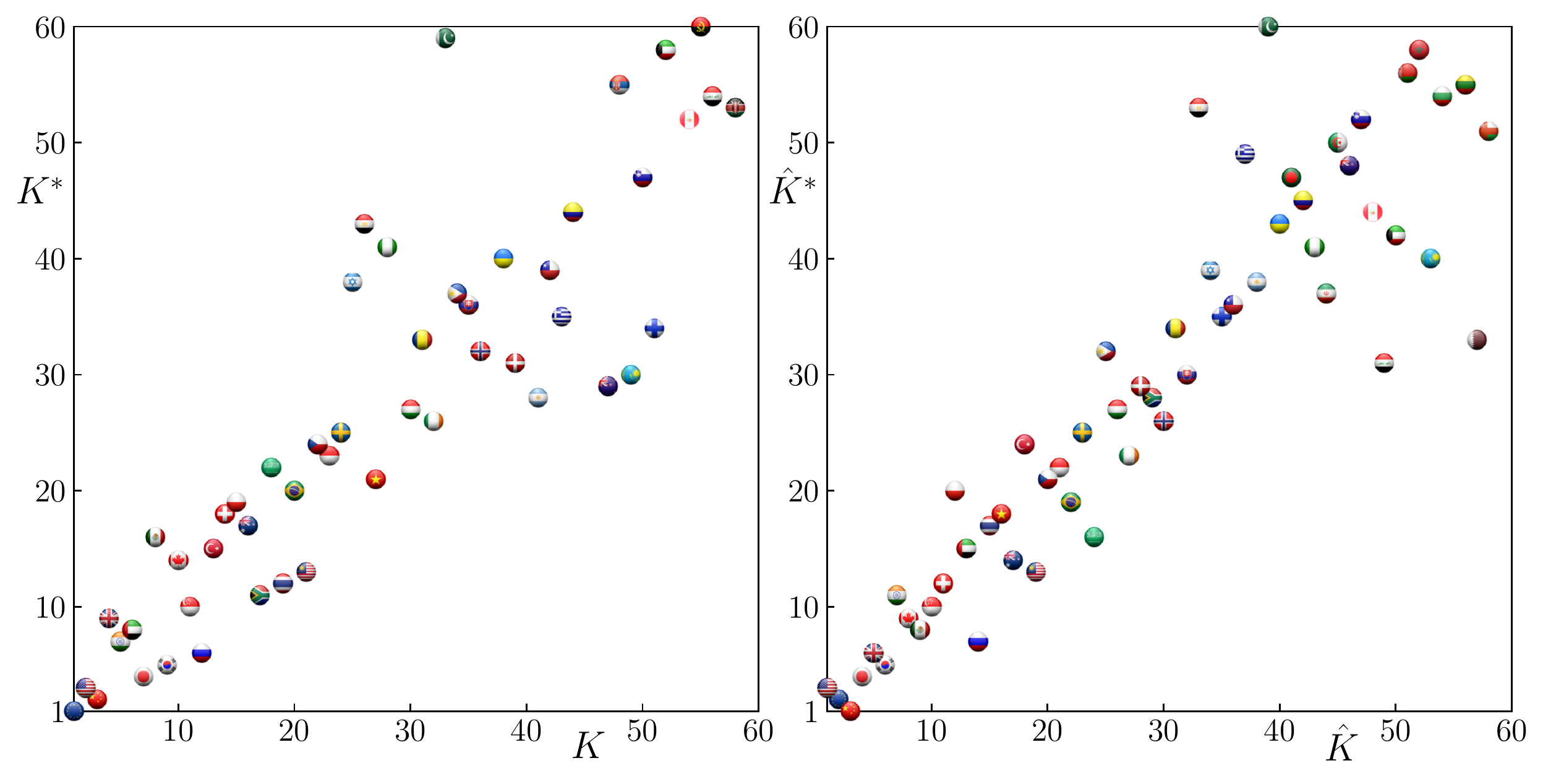}
\caption {\baselineskip 14pt Circles with county flags show positions of countries on
  the plane of PageRank-CheiRank indexes $(K,K^*)$
  (summation is done over all products) (left panel)
  and on the plane of ImportRank-ExportRank  $\hat{K}$, $\hat{K}^*$ from trade volume (right panel);
  data are shown only for index values less than $61$ for year 2018;
  data for years 2012, 2014, 2016  are shown in SupInf Fig.S1; KEU9 is marked by EU flag.}
\label{fig1}\label{figure1}
\end{center}
\end{figure*}

We start the presentation of obtained results
from showing the distribution
of world countries on the plane
of CheiRank-PageRank indexes $(K,K^*)$
given in Fig.~\ref{fig1} (left panel).
Here, for a better visibility, we show only countries with $K,K^* \leq 60$,
each country is marked by a circle with its flag.
For a comparison we also present in Fig.~\ref{fig1} (right panel)
the distribution of countries 
on the plane of ExportRank-ImportRank $\hat{K}$, $\hat{K}^*$
(in both panels, for compactness, we keep index $K$ which in fact
corresponds to $K_c$ index of a country obtained
by a summation over all products).
The top 20 countries with their indexes
are given in Table~\ref{tab2}.

The main feature of Fig.~\ref{fig1} and Table~\ref{tab2}
is that KEU9 takes the top leading position in 
PageRank and CheiRank indexes $K, K^*$ in 2018
(this leadership is also present in other studied
years 2012, 2014, 206 as it is shown in Supporting Information (SupInf) Fig.S1).
This result is significantly different from
the Import-Export volume ranking where in 2018
China is leading in export and USA in import.
We argue that the Google matrix analysis
via PageRank and CheiRank treats  in a deeper way
the multiplicity of trade relations between world
countries compared to the standard Import-Export approach
which takes into account only one step trade links.

Another important feature of Google matrix analysis
is a significant improvement of positions of
certain countries compared to their 
usual Import-Export ranking (see Fig.~\ref{fig1}, Table~\ref{tab2}).
Thus Russia moves to the fourth CheiRank position $K^*=6$ compared to its 
ExportRank  $\hat{K}^* =7$. Also India has strong
CheiRank-PageRank position $K^*=7, K=5$
compared to Export-ImportRanks  $\hat{K}^* =11$, $\hat{K}= 7$.
Also there is a significant reduction of positions of Switzerland
from   $\hat{K}^* =12$, $\hat{K}= 11$ to
$K^*=18, K=14$.
In our opinion these results demonstrate 
a significant hidden power or weakness of
trade relations of certain countries
due to the multiplicity and variety
of their trade relations 
which are not visible in a standard  Export-Import approach.

The main message of the results of this part
is the world top leading position of KEU9
in CheiRank and PageRank trade that gives 
confirmation of the strength and importance of
KEU9 countries discussed in \cite{stetienne}.

\subsection*{Trade balance of countries}

\begin{figure*}[!ht]
\begin{center}
\includegraphics[width=0.92\columnwidth,angle=0]{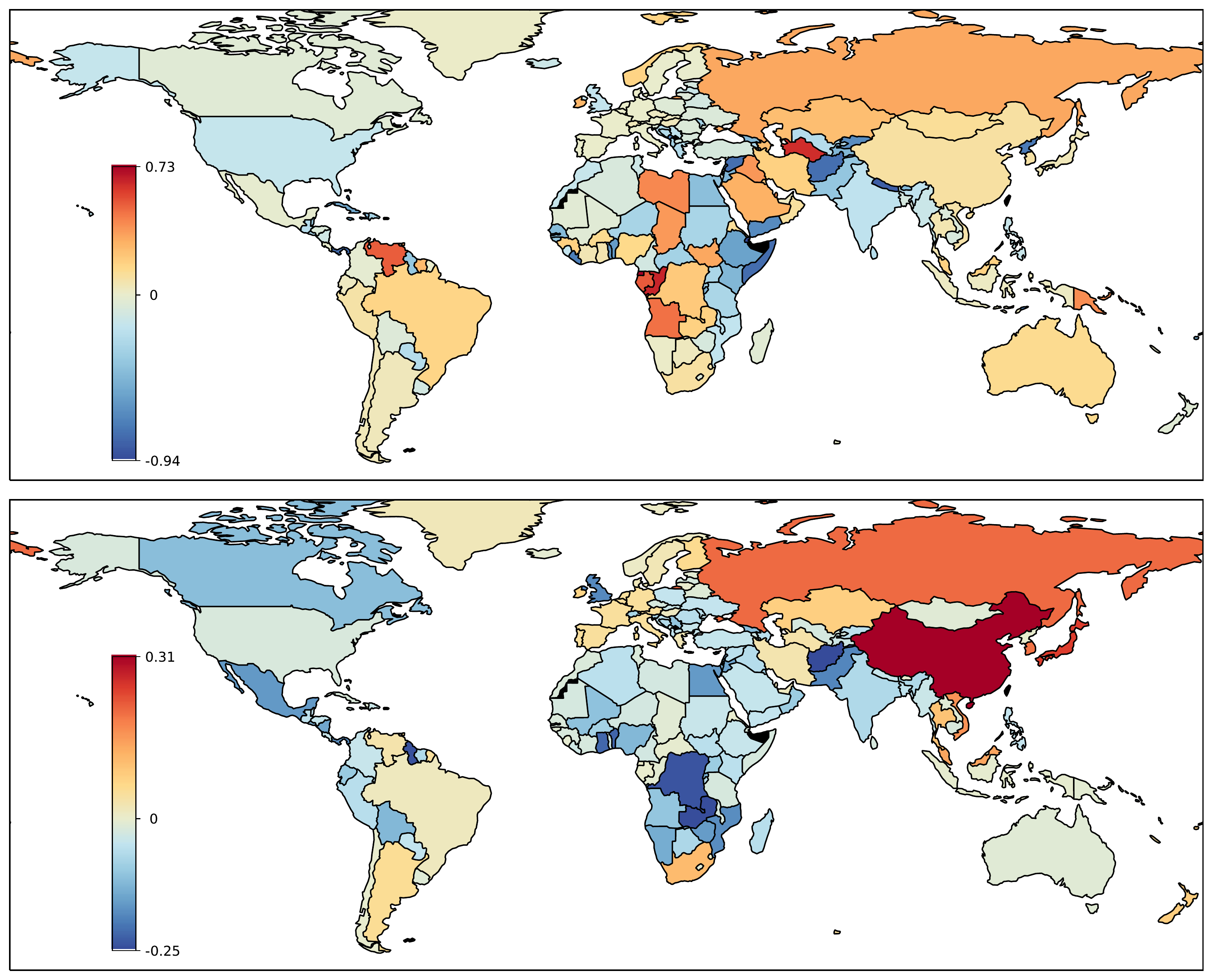}
\caption {\baselineskip 14pt World map of probabilities balance 
$B_c = ({P_c}^* - P_c)/({P_c}^* + P_c)$ of countries.
Top panel: probabilities are computed from the trade volume of Export-Import.
Bottom panel: probabilities are computed from PageRank
and CheiRank vectors.
Data are shown for year 2018 (similar data for years 2012, 2014, 2016 are given in
SupInf Fig.S2). Balance values are marked by color with the corresponding 
color bar ($j$); countries absent in the UN COMTRADE report are marked by black color 
(here and in other Figs.)
}
\label{fig2}\label{figure2}
\end{center}
\end{figure*}

We present the world map of trade balance $B_c$ of countries
obtained from CheiRank-PageRank and ExportRank-ImportRank
probabilities in Fig.~\ref{fig2} for year 2018
(other years 2012, 2014, 2016 are given in SupInf FigS2; 
the distributions of import and export of countries 
for all years are shown in Figs.S3,S4). 
The comparison of two ways of
balance computation
shows that Export-Import approach
does not capture the influence of Russia and China on the world 
trade exchange. In contrast the CheiRank-PageRank  approach
directly highlights the multistep network influence of Russia
and China on the world trade flows and their balance.
We also see a strong positive CheiRank-PageRank balance
for Japan. 
In both approaches the balance of US is close slightly negative.
There is a relative increase of KEU9 balance in CheiRank-PageRank
description compared to the standard  Export-Import one.
We attribute this to the fact that  CheiRank-PageRank description 
takes into account
the multiplicity of trade links which better describes 
a broad variety of KEU9 trade exchange.

We note that by definition we have the balance bounds 
$-1 \leq B_c \leq 1$. The actual obtained bounds are 
$(-0.94,0.73)$ and $(-0.25,0.31)$ for 
Export-Import and CheiRank-PageRank descriptions (see Fig.~\ref{fig2}).
We attribute a reduction of bounds for the latter case
to a multiplicity of network links that reduce
fluctuations in trade exchange.

\subsection*{Sensitivity of trade balance to specific products}

\begin{figure*}[!ht]
\begin{center}
\includegraphics[width=0.92\columnwidth,angle=0]{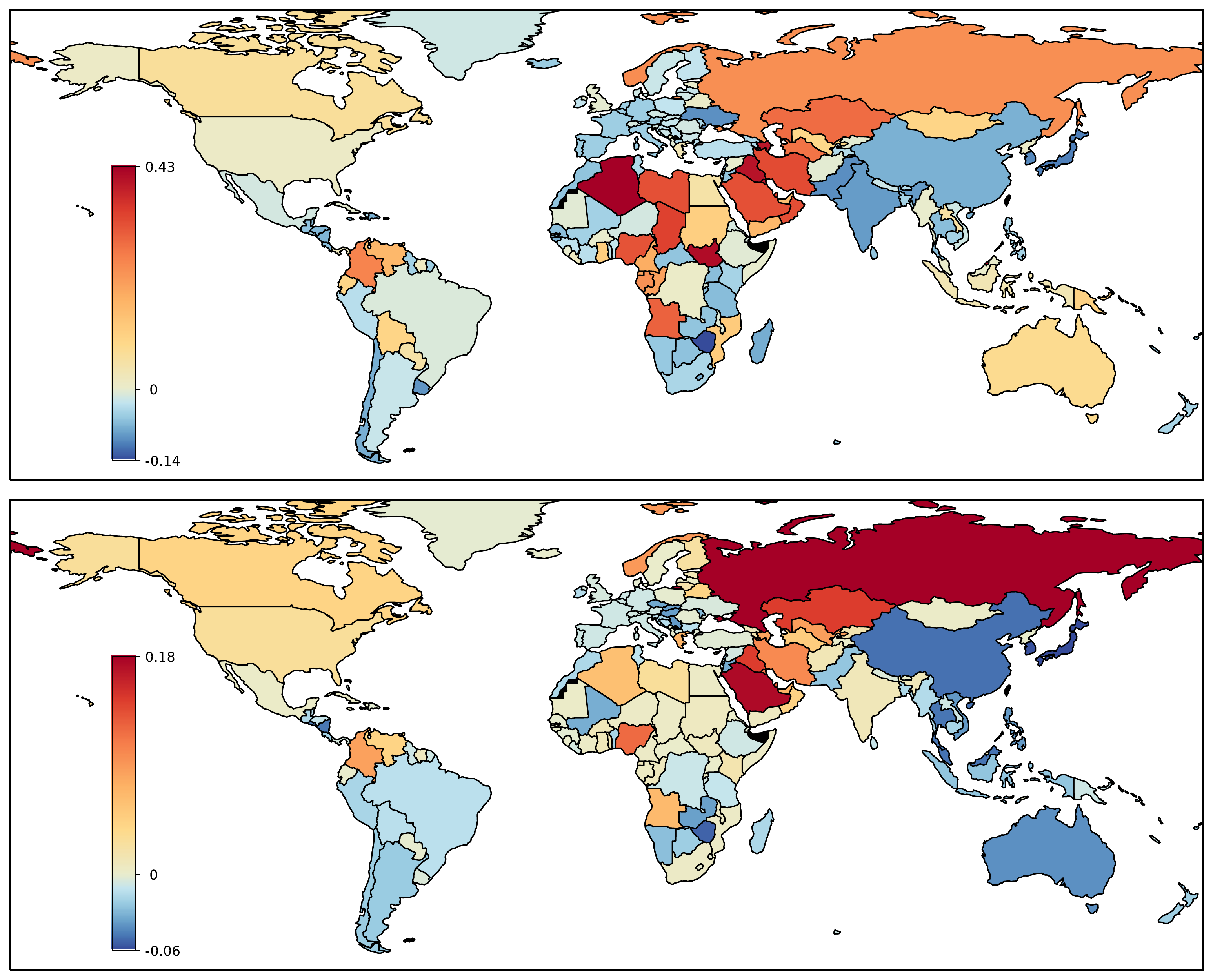}
\caption {\baselineskip 14pt Sensitivity of country balance 
$dB_c/d\delta_s$ for product $s=3$ (mineral fuels).
Top panel: probabilities are computed from the trade volume of Export-Import.
Bottom panel: probabilities are computed from PageRank
and CheiRank vectors. Sensitivity are marked by color with the corresponding 
color bar ($j$). Data are shown for year 2018} 
\label{fig3}\label{figure3}
\end{center}
\end{figure*}

As described above we determine the sensitivity
of trade balance of countries $d B_c/d \delta_s$ to
specific products using the sensitivity
definition from CheiRank-PageRank and Export-Import probabilities.
The sensitivity results for $s=3$ (mineral fuels) are
given in Fig.~\ref{fig3}. The CheiRank-PageRank approach
shows that the most profitable countries 
with the highest values of $d B_c/d \delta_3$
are Saudi Arabia and Russia (Kazakhstan also
has high sensitivity). 
This is rather natural since these countries 
are the highest petroleum producers.
The strongly negative impact is well visible for
Australia, China and countries of Latin America.
USA and KEU9 sensitivities being close to zero.

In contrast the sensitivity from Export-Import approach
gives of the top position Algeria (followed by Brunei).
Among countries with strongly negative
sensitivities we have India, Pakistan and China
while Australia is slightly positive.
In this Export-Import approach USA is slightly positive and 
KEU9 is slightly negative.

This shows a significant difference between
the usual Export-Import analysis and
the Google matrix approach. We argue that the 
latter approach takes into account the multiplicity of 
trade links and flows thus highlighting
in a better way the multistep trade relations between countries.

\begin{figure*}[!ht]
\begin{center}
\includegraphics[width=0.92\columnwidth,angle=0]{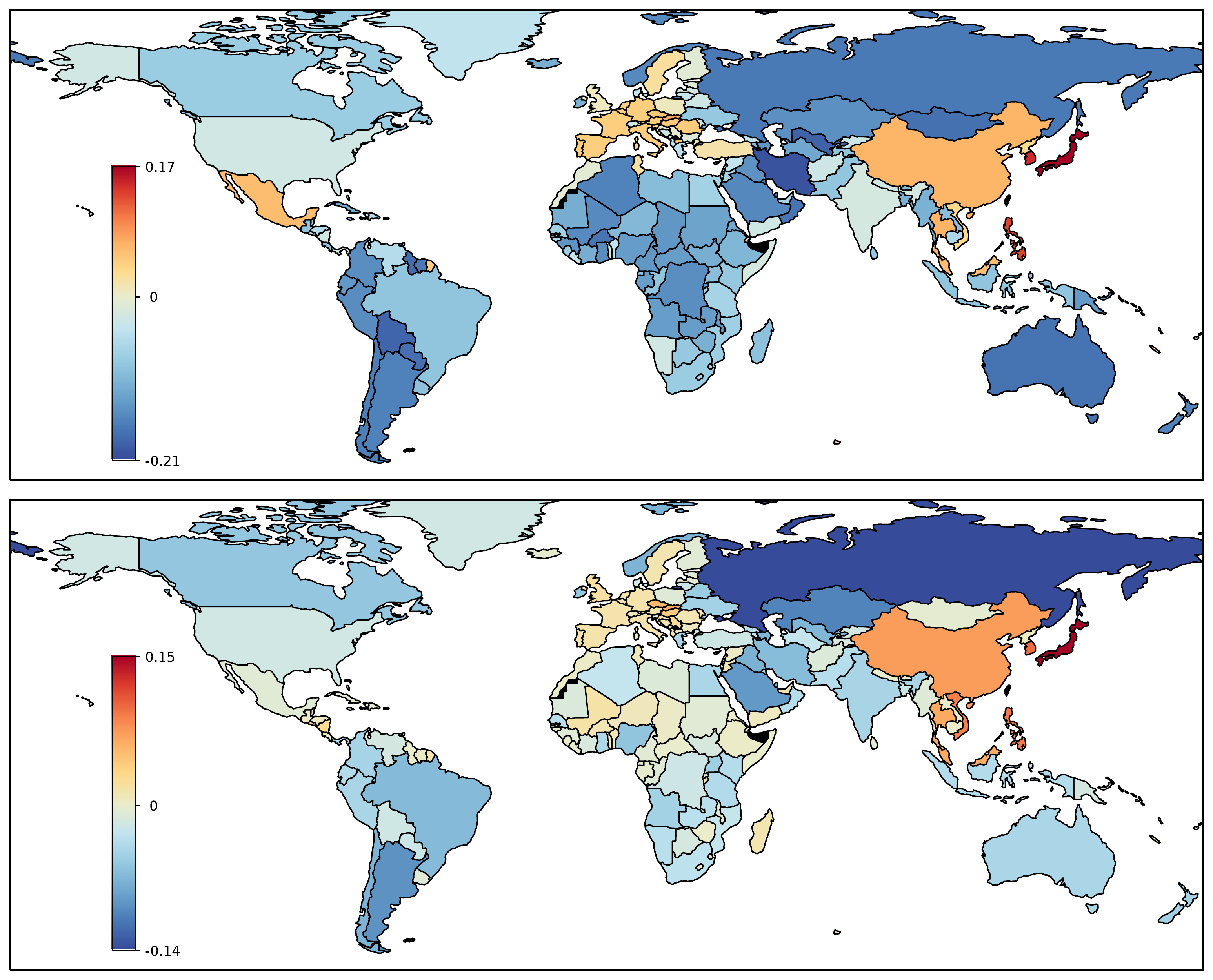}
\caption {\baselineskip 14pt Same as in Fig.~\ref{fig3} but
for product $s=7$ (machinery). Data are shown for year 2018
 (similar data for years 2012, 2014, 2016 are given in
SupInf Fig.S6).}
\label{fig4}\label{figure4}
\end{center}
\end{figure*}

The sensitivities of countries to the product $s=7$ (machinery)
is shown in Fig.~\ref{fig4}. Here both approaches 
give the most positive countries
being Japan, S.Korea and China. In the Export-Import approach KEU9
has a bit higher positive sensitivity
compared to the CheiRank-PageRank method.
Thus we have 
for both methods of CheiRank-PageRank and Export-Import: 
KEU9 $dB_c/d\delta_7 = 0.015$, $d\hat{B}_c/d\delta_7 = 0.043$;
slightly negative values for USA $dB_c/d\delta_7 = -0.019$, $d\hat{B}_c/d\delta_7 = -0.027$;
Russia has strongly negative values $dB_c/d\delta_7 = -0.145$, $d\hat{B}_c/d\delta_7 = -0.169$.

In the above Figs.~\ref{fig3},~\ref{fig4} we presented results
for year 2018. The same type of data for years 2012, 2014, 2016
are given in SupInf Figs.~S5,S6.

\begin{figure*}[!ht]
\begin{center}
\includegraphics[width=0.92\columnwidth,angle=0]{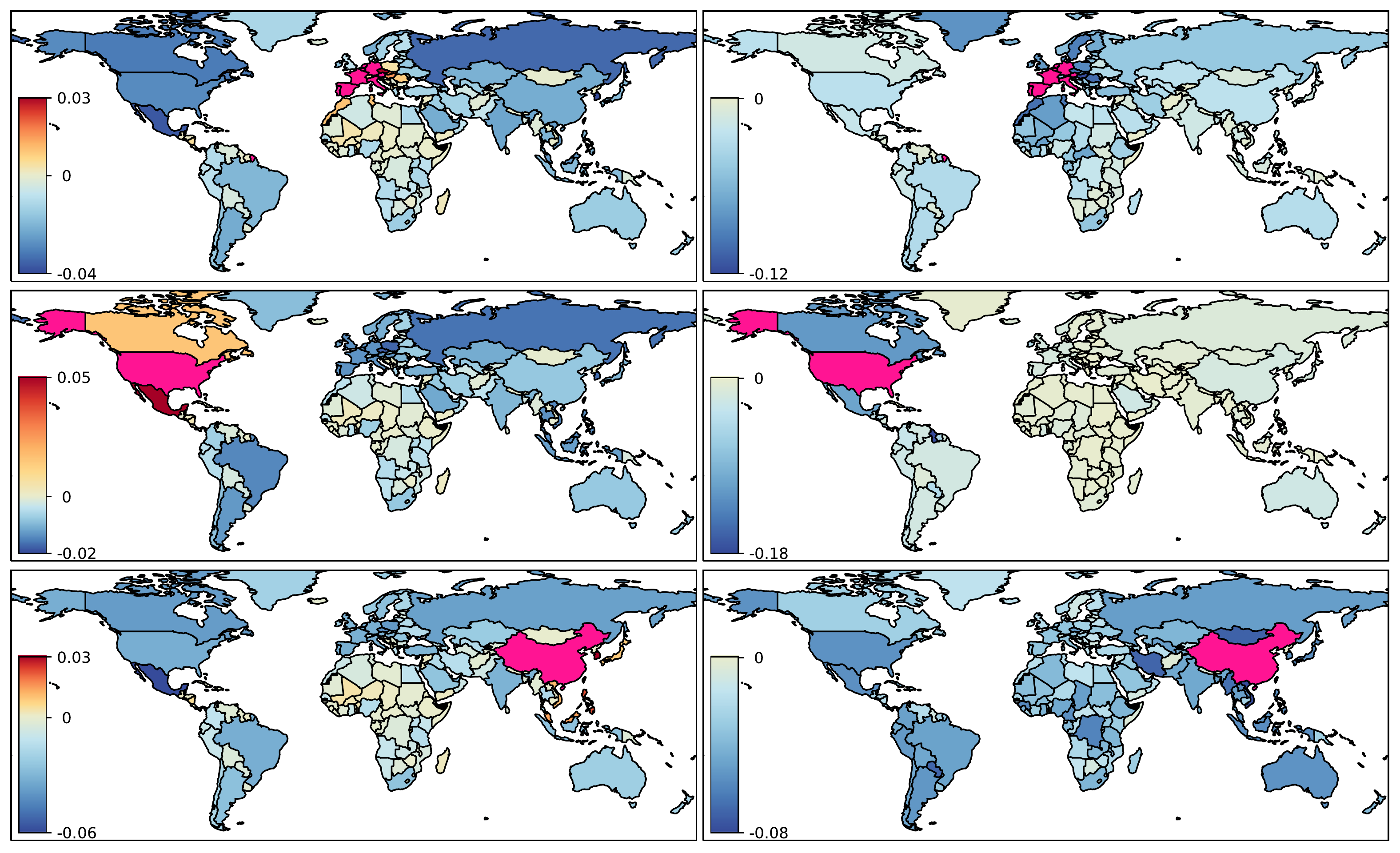}
\caption {\baselineskip 14pt Sensitivity of country balance 
$dB_c/d\delta_{cs}$ for product price $s=7$ (machinery)
of KEU9 (top row), USA (middle row), China (bottom row).
Left column: probabilities are computed from PageRank
and CheiRank vectors.
Right column: probabilities are computed from the trade volume of Export-Import.
Data are shown for year 2018.
Balance values are marked by color with the corresponding 
color bar ($j$). Sensitivity values for
KEU9, USA, China are  $dB_c/d\delta_{cs} = 0.12, 0.12, 0.15$ (left column)
and $0.21, 0.18, 0.24$ (right column) respectively, these values are 
marked by separate magenta color to highlight sensitivity of other countries in a better way.} 
\label{fig5}\label{figure5}
\end{center}
\end{figure*}

Above we considered the sensitivity of trade balance to a 
global price variation of a given product applicable to 
the whole world with a homogeneous price increase of product for all countries.
It is also interesting to consider the sensitivity
of country trade balance when the product price is changed only by one country.
In this way we obtain the sensitivity $d B_c/ d \delta_{cs}$ 
of countries to a product of a given country.
This specific  sensitivity  is shown in Fig.~\ref{fig5}
in respect to price variation of $s=7$ (machinery)
from KEU9, USA, and China in year 2018.
The Export-Import approach gives strongly positive
sensitivity for the country which increased price 
of machinery (respectively KEU9, USA, China).
All other countries have sensitivity close to zero or negative.
The result from CheiRank-PageRank analysis is 
different. For KEU9 machinery price increase
 the  positive sensitivity
is obtained for Czechia, Slovakia, Hungary 
(with the sensitivity values $0.028, 0.017, 0.015$ respectively).
For USA case the  positive sensitivity
is obtained for  Mexico, Canada with repective values $0.045, 0.014$.
For China case the  positive sensitivity
is obtained for Korea, Philippines, Malaysia 
(with sensitivity respective values $0.031,0.023,0.014$).
This increase is related to strong network links between 
these countries well captured by the Google matrix analysis.

\begin{figure*}[!ht]
\begin{center}
\includegraphics[width=0.92\columnwidth,angle=0]{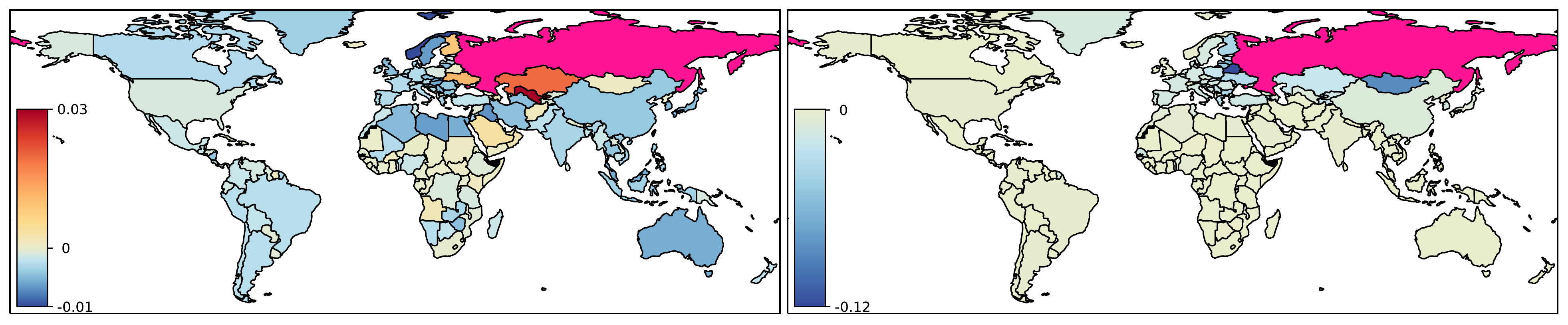}
\caption {\baselineskip 14pt Sensitivity of country balance 
$dB_c/d\delta_{cs}$ for product price $s=3$ (mineral fuels)
of Russia.
Left panel: probabilities are computed from PageRank
and CheiRank vectors; 
Right panel: probabilities are computed from the trade volume of Export-Import.
Data are shown for year 2018.
Sensitivity values for Russia are  $dB_c/d\delta_{cs} = 0.13$ (left panel)
and $0.24$ (right panel);  these values are 
marked by separate magenta color to highlight sensitivity of other countries in a better way.}
\label{fig6}\label{figure6}
\end{center}
\end{figure*}

In Fig.~\ref{fig6} we show the sensitivity $d B_c/ d \delta_{cs}$ from both approaches
for $s=3$ (mineral fuels) of Russia. Again as for $s=7$ we
see that the Export-Import approach gives
the strong positive sensitivity only for Russia.
In contrast the CheiRank-PageRank approach 
shows that  Uzbekistan, Kazakhstan, Ukraine 
(with values $d B_c/ d \delta_{cs} = 0.032,0.021,0.012$ respectively)
also gain the positive sensitivity in the case of price increase of $s=3$ of Russia.
This also confirms the strength of Google matrix analysis
which captures multiple trade links between countries.

\subsection*{Sensitivity of trade balance to labor cost}

It is interesting to analyze the sensitivity 
of a country trade balance $d B_c/d \sigma_{c'}$ to
a labor cost variation in a given country.
This analysis is done by  increasing 
the price of all products of a given
country by a factor $1+\sigma_{c'}$
followed by a renormalization of sum all column elements to unity.
Such an approach has been developed and studied in  \cite{escaith}
for the world economic activities from World Trade Organization data.
Here, at the difference of price shock of one product,
the price increase affects all product flows from a given country
corresponding to a global increase of labor
cost in a given country. Of course, the price increase
is considered to be very small corresponding to the linear response
regime. The labor cost sensitivity $d B_c/d \sigma_{c'}$ is computed numerically
in the same manner as the product sensitivity $d B_c/d \delta_{c'}$
discussed above. 

As discussed in \cite{escaith}
the most strong labor cost sensitivity  $d B_c/d \delta_{c'}$
is naturally obtained for the country itself with $c=c'$. 
Therefore, below in Fig.~\ref{fig7} we present the diagonal results
for $d B_c/d \delta_{c'}$ at $c \neq c'$ by a separate magenta color
while all other countries sensitivity are characterized by
color bar. 

For KEU9 the strongest sensitivity values $d B_c/d \delta_{c'}$ are obtained for 
countries: Czechia, Tunisia, Morocco with positive values $0.027, 0.014,0.013$
and S.Korea, Canada, Russia with negative values $-0.074,-0.072,-0.062$.
For USA case these are: Canada, Mexico, Venezuela with positive 
values  $0.086,0.063,0.008$ and  S.Korea, Japan, United Kingdom with negative 
values $-0.045,-0.044,-0.044$;
for China we find:  Korea, Congo, Philippines with positive
values $0.033,0.015,0.014$ and  Mexico, Canada, Poland with negative values
$-0.090, -0.076, -0.074$.
For Russia  we obtain: Uzbekistan, Belarus, Kazakhstan 
with positive values $0.028, 0.026,0.020$ and Norway, Sweden, United Kingdom
with negative values $-0.019,-0.016,-0.013$.

This shows that the Google matrix analysis allows to determine
on pure mathematical grounds the mutual trade dependences of
world countries.

\subsection*{Network structure of trade from reduced Google matrix}

We use the REGOMAX algorithm described above to
obtain the reduced Google matrix of trade flows between certain selected
countries. We choose the case of 4 countries which has a strong world 
trade influence KEU9, USA, China, Russia 
with 10 trade products of Table~\ref{tab1}.
In this way the size of $G_R$ (Import or PageRank direct flow direction) 
and ${G^*}_R$ (Export or CheiRank inverted flow direction) is equal to $40$.
For clarity we show only 4 directed links corresponding to the most strong 
matrix elements from a given node (only non-diagonal terms are shown).
The obtained networks are shown in Fig.~\ref{fig8} and Fig.~\ref{fig9}
respectively.

The network structure shown in Fig.~\ref{fig8} from  $G_R$ 
shows that the main importing nodes are machinery ($s=7$)
of KEU9 and USA. In a similar way the main exporting nodes of ${G^*}_R$
are again machinery product of KEU9, China and USA (Fig.~\ref{fig9}).
This clearly show the importance of machinery product
for the world trade.

\section*{Discussion}

Above we considered the trade influence of kernel EU 9 countries (KEU9) 
considered as a one united state following the proposal 
pushed forward by Christian Saint-Etienne in \cite{stetienne}.
The analysis is done on the bases of multiproduct trade data
provided by UN COMTRADE \cite{comtrade}. Our results are based on 
the advanced Google matrix analysis
of multiproduct world trade network flows for years 2012-2018
between all world countries registered at UN.
They clearly show that KEU9 takes the world leading 
position in PageRank and CheiRank probabilities
being ahead of USA and China.
This mathematical network analysis
demonstrates that KEU9 becomes the main player in the international trade.
This provides additional mathematical foundation for
the historical, economical and political arguments presented 
in  \cite{stetienne}
in the favor of coherent strong impact of KEU9 (if united)
on the world development.

We also show that the Google matrix analysis allows
to obtained significantly deeper information about
world trade comparing to the Import-Export
analysis usually used in economy studies.


\section*{Acknowledgments}
We thank Katia Jaffres-Runser (INP ENSEEIHT Toulouse)
for useful discussions.
This research has been partially supported through the grant
NANOX $N^\circ$ ANR-17-EURE-0009 (project MTDINA) in the frame 
of the {\it Programme des Investissements d'Avenir, France} and
in part by APR 2019 call of University of Toulouse and by 
Region Occitanie (project GoIA).
We thank UN COMTRADE for providing us a friendly access
to their detailed database.

\section*{Supporting Information}

Supporting Information file
presents Figures S1-S6 extending information presented in Figures of the main part to
years 2012, 2014, 2016 and more details.

\newpage$\phantom{.}$

\begin{table*}[!ht]
\caption{Product code and name for SITC classification level 1} 
\begin{center}
\begin{tabular}{|c|c|}
  \hline
Code	&	Name	\\
\hline
0	&	Food and live animals	\\
1	&	Beverages and tobacco	\\
2	&	Crude materials,inedible,except fuels	\\
3	&	Mineral fuels etc	\\
4	&	Animal and vegetable oils and fats	\\
5	&	Chemicals and related products,n.e.s.	\\
6	&	Basic manufactures	\\
7	&	Machinery,transport equipment	\\
8	&	Miscellaneous manufactured articles	\\
9	&	Goods not classified elsewhere	\\
  \hline
\end{tabular}
\end{center}
\label{tab1}
\end{table*}

\begin{table*}[!ht]
\caption{Top 20 Ranking of PageRank ($K$), CheiRank ($K^*$), 
ImportRank and ExportRank for the year 2018. 
The table lists country ranks of both panels of Fig.~\ref{fig1}.}
\begin{center}
\begin{tabular}{|c|c|c|c|c|}
  \hline
Rank	&	PageRank ($K$)	&	CheiRank ($K^*$)	&	ImportRank	&	ExportRank	\\
  \hline
1	&	KEU9	&	KEU9	&	USA	&	China	\\
2	&	USA	&	China	&	KEU9	&	KEU9	\\
3	&	China	&	USA	&	China	&	USA	\\
4	&	United Kingdom	&	Japan	&	Japan	&	Japan	\\
5	&	India	&	Republic of Korea	&	United Kingdom	&	Republic of Korea	\\
6	&	U. Arab Emirates  &	Russian Federation	&	Republic of Korea	&	United Kingdom	\\
7	&	Japan	&	India	&	India	&	Russian Federation	\\
8	&	Mexico	&	U. Arab Emirates	&	Canada	&	Mexico	\\
9	&	Republic of Korea	&	United Kingdom	&	Mexico	&	Canada	\\
10	&	Canada	&	Singapore	&	Singapore	&	Singapore	\\
11	&	Singapore	&	South Africa	&	Switzerland	&	India	\\
12	&	Russian Federation	&	Thailand	&	Poland	&	Switzerland	\\
13	&	Turkey	&	Malaysia	&	U. Arab Emirates	&	Malaysia	\\
14	&	Switzerland	&	Canada	&	Russian Federation	&	Australia	\\
15	&	Poland	&	Turkey	&	Thailand	&	U. Arab Emirates	\\
16	&	Australia	&	Mexico	&	Vietnam	&	Saudi Arabia	\\
17	&	South Africa	&	Australia	&	Australia	&	Thailand	\\
18	&	Saudi Arabia	&	Switzerland	&	Turkey	&	Vietnam	\\
19	&	Thailand	&	Poland	&	Malaysia	&	Brazil	\\
20	&	Brazil	&	Brazil	&	Czechia	&	Poland	\\
  \hline
\end{tabular}
\end{center}
\label{tab2}
\end{table*}

\begin{figure*}[!ht]
\begin{center}
\includegraphics[width=0.92\columnwidth,angle=0]{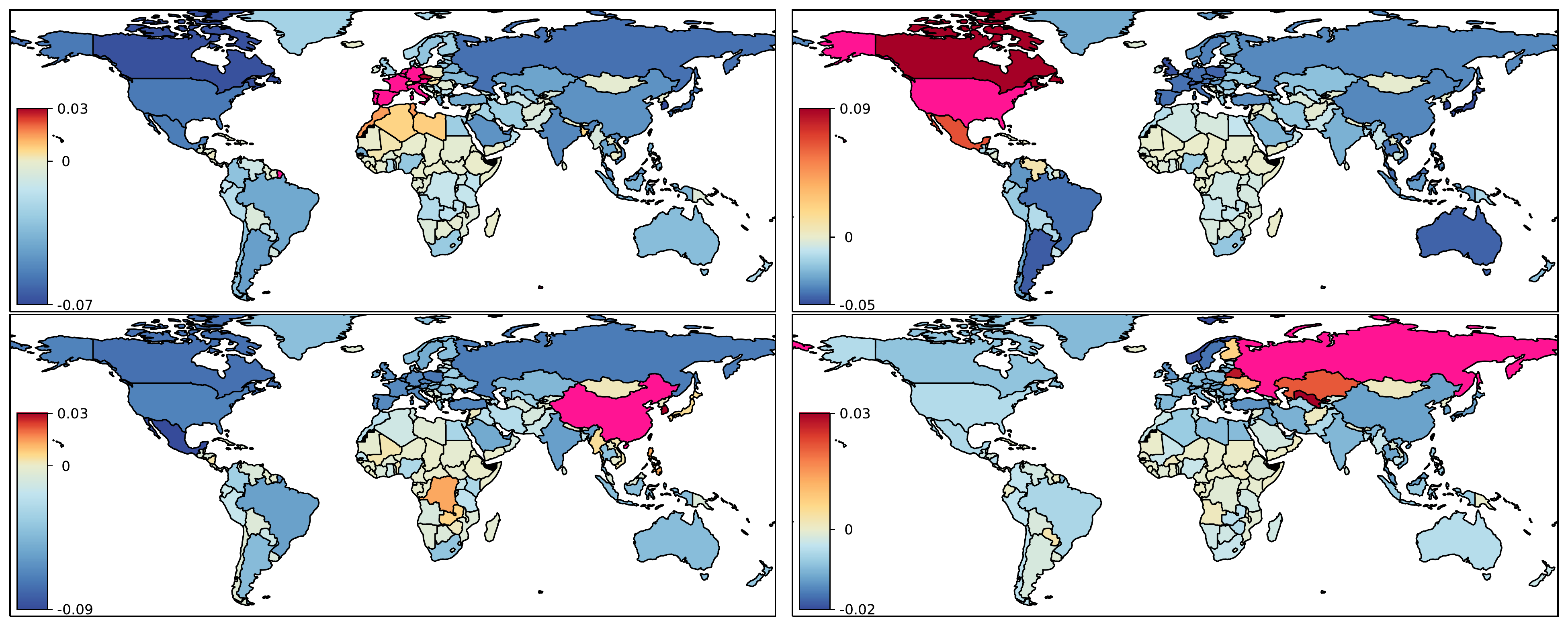}
\caption {\baselineskip 14pt Labor cost sensitivity of country balance 
$dB_c/d\sigma_{c'}$ for KEU9 (top left panel), USA (top right panel), 
China (bottom left panel), Russia (bottom right panel) for year 2018 
(sensitivity is obtained from PageRank-CheiRank probabilities).
The diagonal term $dB_c/d\sigma_{c}$ is shown by a fixed magenta color
with numerical values being $dB_c/d\sigma_{c} = 0.30, , 0.31, 0.30, 0.29$
respectively for KEU9, USA, China, Russia
(this allows to highlight the effect for other countries
in a better way)).}
\label{fig7}\label{figure7}
\end{center}
\end{figure*}

\begin{figure*}[!ht]
\begin{center}
\includegraphics[width=0.92\columnwidth,angle=0]{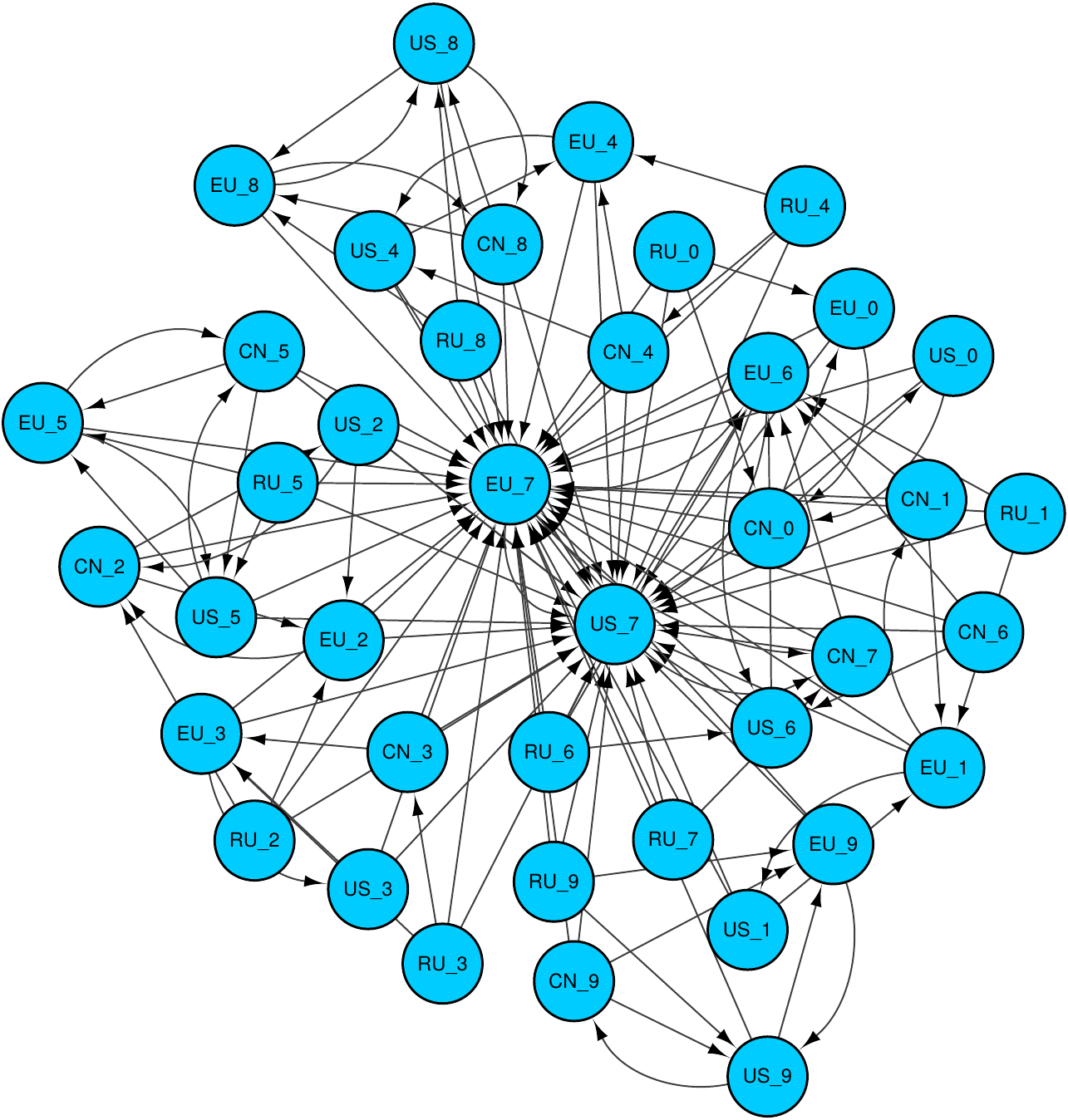}
\caption {\baselineskip 14pt Network trade structure between 
KEU9 (marked as EU), USA (US), China (CN), Russia (RU), Japan (JP)
with 10 products of Table~\ref{tab1} in year 2018. 
Network is obtained from the reduced
Google matrix $G_R$ by tracing four strongest outgoing links.
Countries are shown by circles with two letters of country
and product index from Table~\ref{tab1}. The arrow direction 
from node $A$ to node $B$ means that $B$ imports from $A$.
All $40$ nodes are shown.}
\label{fig8}\label{figure8}
\end{center}
\end{figure*}

\begin{figure*}[!ht]
\begin{center}
\includegraphics[width=0.92\columnwidth,angle=0]{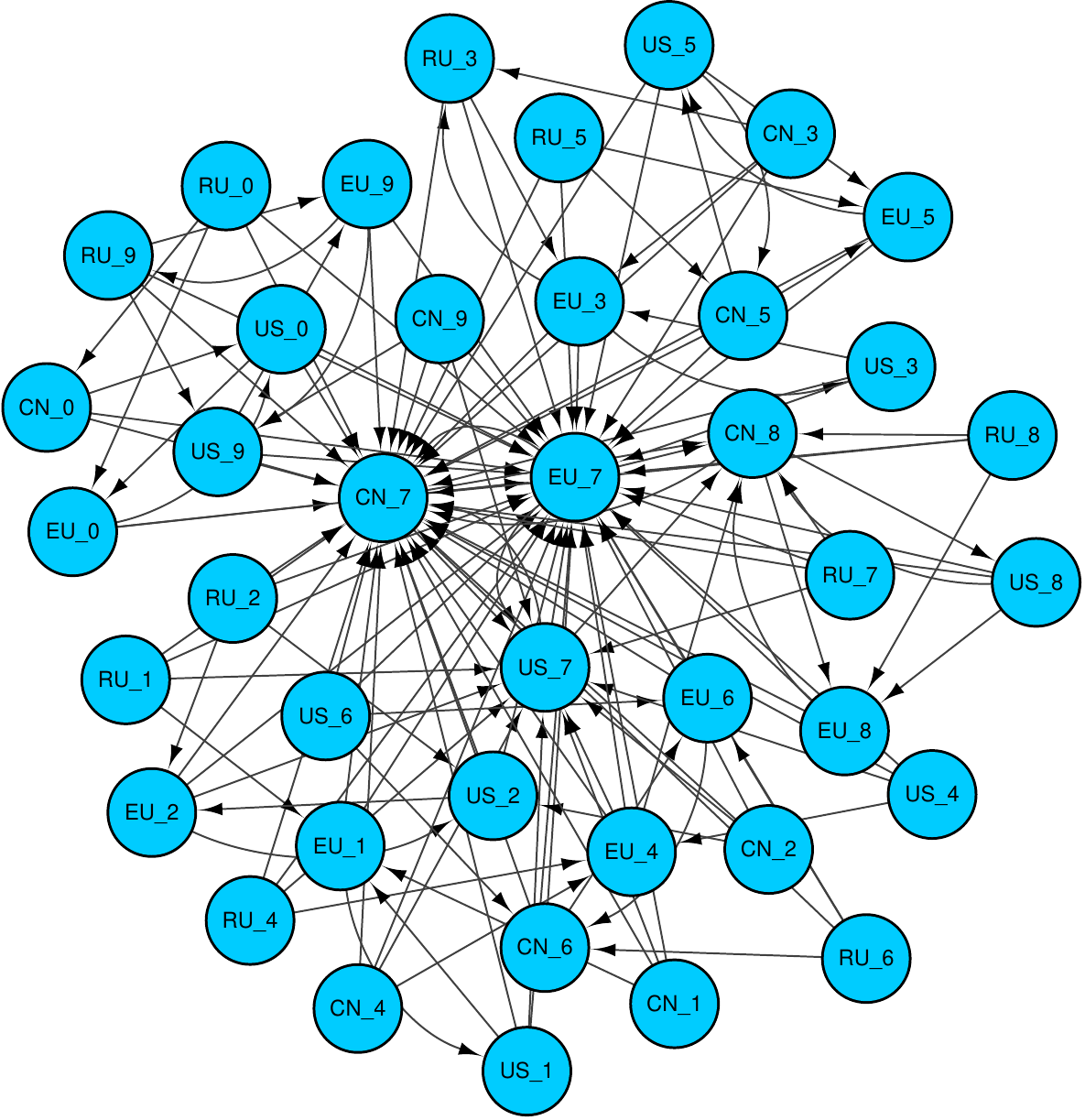}
\caption {\baselineskip 14pt Same in Fig.~\ref{fig8}
but for the reduced Google matrix ${G^*}_R$
corresponding of inverted trade flows (CheiRank or Export direction).
The arrow direction 
from node $A$ to node $B$ means that $B$ exports to $A$.
}
\label{fig9}\label{figure9}
\end{center}
\end{figure*}

\newpage$\phantom{.}$
\FloatBarrier
\begin{flushleft}
  {\Large \textbf{SUPPORTING INFORMATION FOR:\\
World impact of kernel EU 9 countries \\
     from  Google matrix analysis of  UN COMTRADE network} }
\\ \bigskip
Justin Loye$^{1,2}$,
Leonardo Ermann$^{3,4}$,
Dima L.\ Shepelyansky$^{1,*}$
\\ \medskip
{1} \newblock { Laboratoire de Physique Th\'eorique du CNRS, IRSAMC,
Universit\'e de Toulouse, CNRS, UPS, 31062 Toulouse, France}
\\
{2} \newblock { Institut de Recherche en Informatique de Toulouse, 
		Universit\'e de Toulouse, UPS, 31062 Toulouse, France}
\\
{3} \newblock { Departamento de F\'{\i}sica Te\'orica, GIyA,
 Comisi\'on Nacional de Energ\'{\i}a At\'omica.
 Av.~del Libertador 8250, 1429 Buenos Aires, Argentina}
\\
{4} \newblock {  Consejo Nacional de Investigaciones
Cient\'ificas y T\'ecnicas (CONICET), Buenos Aires, Argentina}
\\
\medskip
$\ast$ Webpage: www.quantware.ups-tlse.fr/dima
\end{flushleft}


\section*{Additional data}

Here we present additional Figures related to the material
presented in the main part of the article.

\begin{figure*}[!ht]
\renewcommand\thefigure{S1}
\begin{center}
\includegraphics[width=0.9\columnwidth,angle=0]{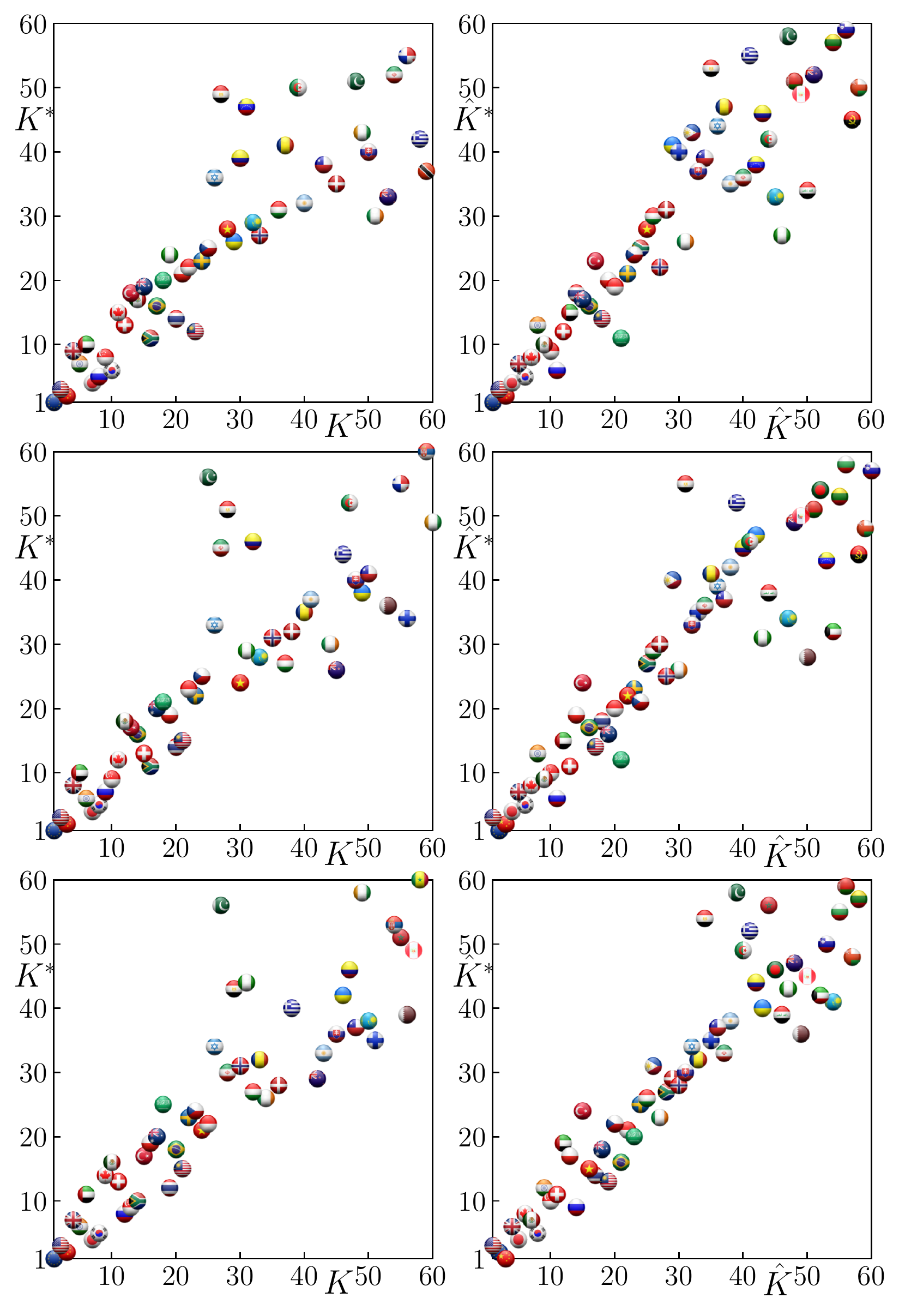}
\caption {\baselineskip 14pt Same as Fig.1 
for years 2012, 2014, 2016 (from top to bottom).}
\label{figS1}\label{figureS1}
\end{center}
\end{figure*}

\begin{figure*}[!ht]
\renewcommand\thefigure{S2}
\begin{center}
\includegraphics[width=0.9\columnwidth,angle=0]{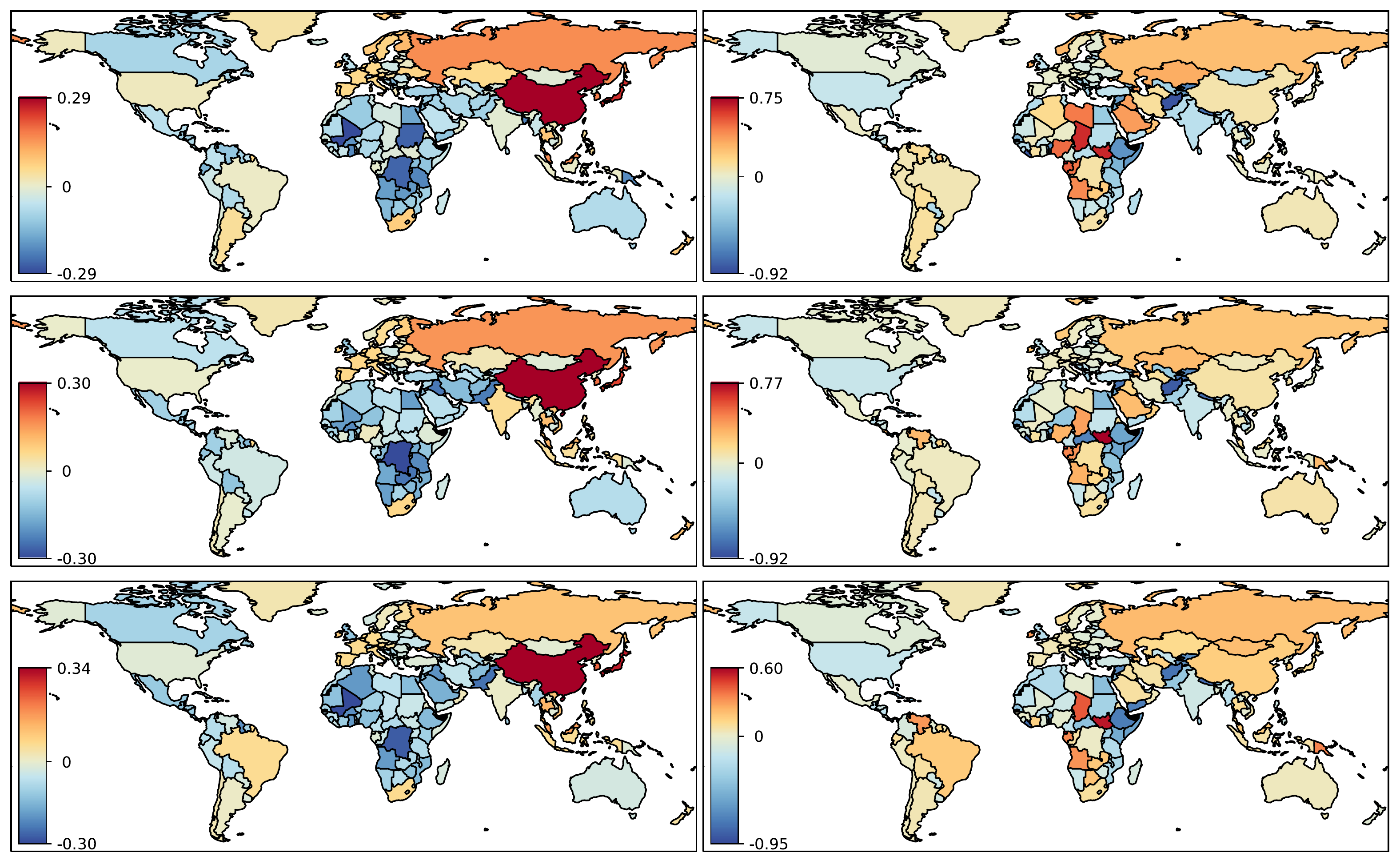}
\caption {\baselineskip 14pt Same as Fig.2, left column
shows data from CheiRank and PageRank vectors,
right column shows data from Export and Import volumes;
data are given 
for years 2012, 2014, 2016 (from top to bottom).}
\label{figS2}\label{figureS2}
\end{center}
\end{figure*}

\begin{figure*}[!ht]
\renewcommand\thefigure{S3}
\begin{center}
\includegraphics[width=0.9\columnwidth,angle=0]{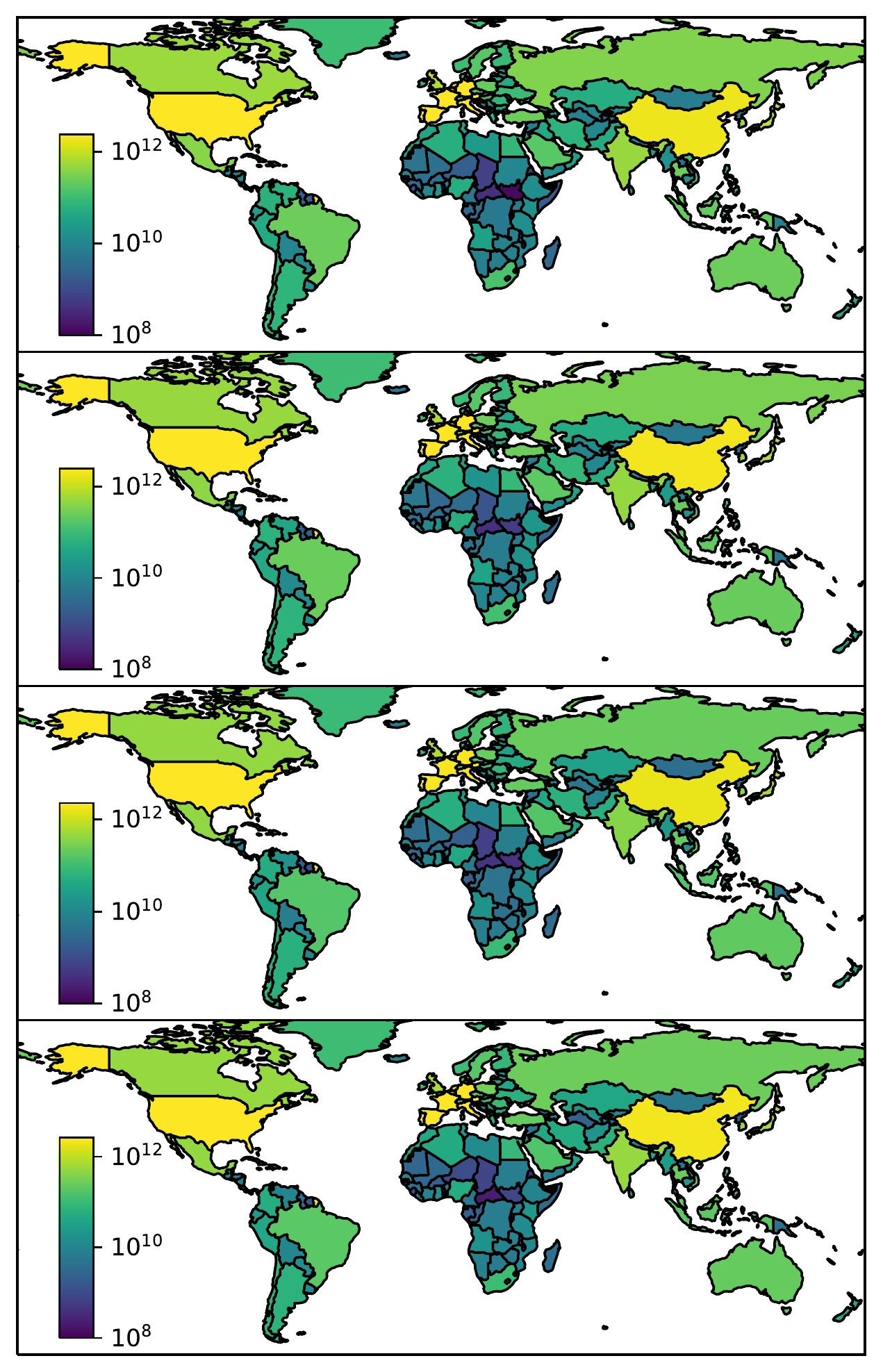}
\caption {\baselineskip 14pt World map of
 countries Import volume in USD
for years 2012, 2014, 2016 (from top to bottom);
values lower then $10^8$ are clipped for better visibility.}
\label{figS3}\label{figureS3}
\end{center}
\end{figure*}

\begin{figure*}[!ht]
\renewcommand\thefigure{S4}
\begin{center}
\includegraphics[width=0.9\columnwidth,angle=0]{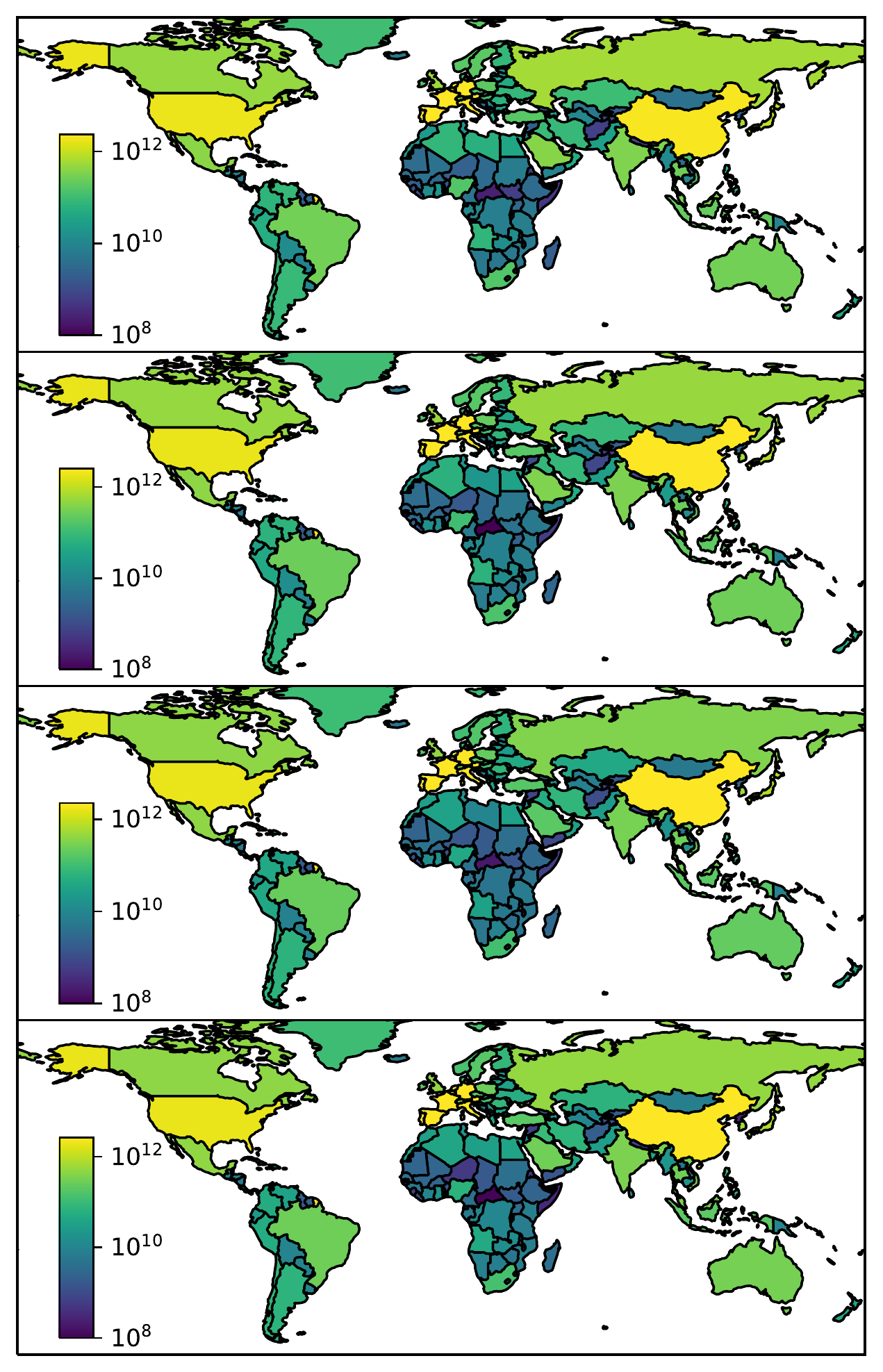}
\caption {\baselineskip 14pt World map of
 countries Export volume in USD
for years 2012, 2014, 2016 (from top to bottom);
values lower then $10^8$ are clipped for better visibility.}
\label{figS4}\label{figureS4}
\end{center}
\end{figure*}

\begin{figure*}[!ht]
\renewcommand\thefigure{S5}
\begin{center}
\includegraphics[width=0.9\columnwidth,angle=0]{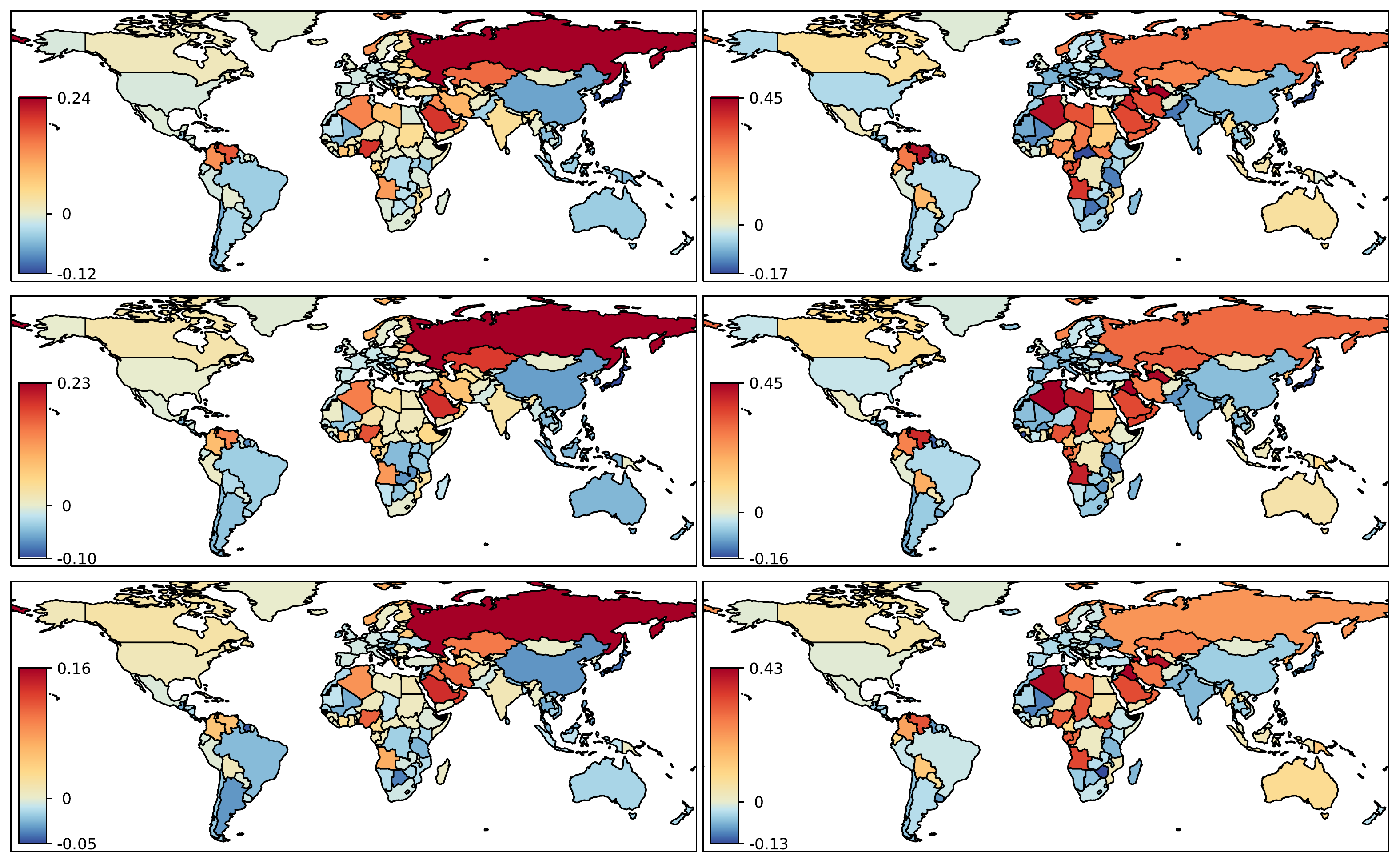}
\caption {\baselineskip 14pt  Same as Fig.3, left column
shows data from CheiRank and PageRank vectors,
right column shows data from Export and Import volumes;
data are given 
for years 2012, 2014, 2016 (from top to bottom).}
\label{figS5}\label{figureS5}
\end{center}
\end{figure*}

\begin{figure*}[!ht]
\renewcommand\thefigure{S6}
\begin{center}
\includegraphics[width=0.9\columnwidth,angle=0]{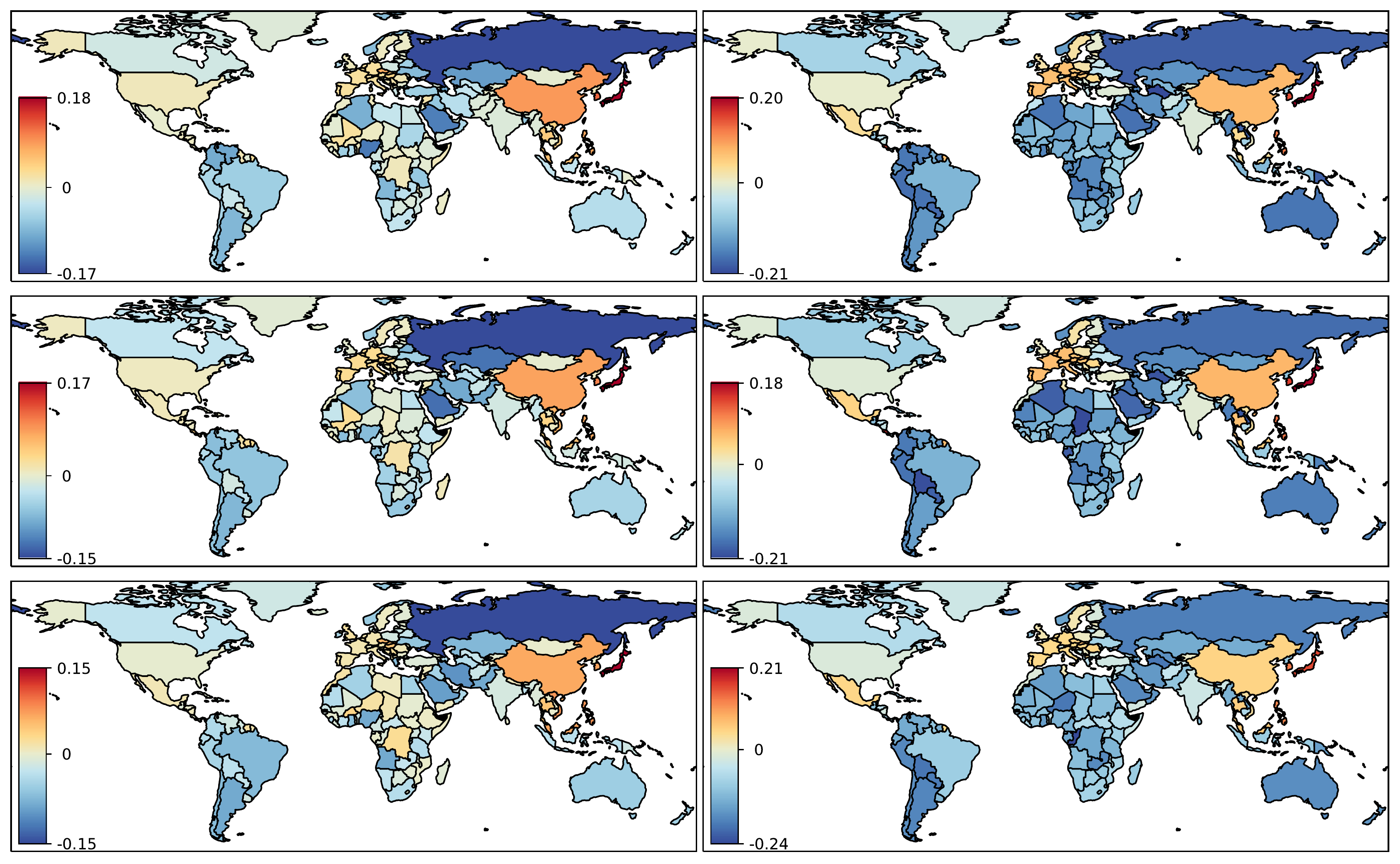}
\caption {\baselineskip 14pt  Same as Fig.4, left column
shows data from CheiRank and PageRank vectors,
right column shows data from Export and Import volumes;
data are given 
for years 2012, 2014, 2016 (from top to bottom).}
\label{figS6}\label{figureS6}
\end{center}
\end{figure*}


\begin{thebibliography}{99}

\baselineskip 15pt

\bibitem{stetienne} 
C.~Saint-Etienne 
\newblock (2018)
\newblock {Osons l'Europe des nations}
\newblock Editions de l'Observatoire/Humensis 

\bibitem{wikieconomy} 
Wikipedia:{\it Economy of the European Union}\\
\newblock https://en.wikipedia.org/wiki/Economy\_of\_the\_European\_Union
(Retrieved  October, 2020)

\bibitem{china1st} 
Largest Economies in the World\\
\newblock https://www.thebalance.com/world-s-largest-economy-3306044 
(Retrieved October, 2020)

\bibitem{wikieu} 
Wikipedia: {\it European Union} \\
\newblock https://en.wikipedia.org/wiki/European\_Union
(Retrieved October, 2020)

\bibitem{wikieupolit} 
Wikipedia: {\it Politics of the European Union}\\
\newblock https://en.wikipedia.org/wiki/Politics\_of\_the\_European\_Union
(Retrieved October, 2020)

\bibitem{zimour} 
{\it CNEWS: Face a l'info avec Eric  Zemmour Saison 1, 14 Nov 2019}\\
\newblock https://www.cnews.fr/emission/2019-11-14/face-linfo-du-14112019-899420
(Retrieved October 2020)

\bibitem{comtrade} 
United Nations Commodity Trade Statistics Database\\
\newblock Available: http://comtrade.un.org/db/. 
(Retrieved  October 2020).

\bibitem{wto2018} 
World Trade Organization (2018)
\newblock { World Trade Statistical Review 2018}\\
Available: https://www.wto.org/english/res\_e/statis\_e/wts2018\_e/wts18\_toc\_e.htm
(Retrieved October 2020).

\bibitem{krugman2011} 
Krugman PR, Obstfeld M, Melitz M
\newblock (2011)
\newblock {International economics: theory \& policy}
\newblock Prentic Hall, New Jersey 

\bibitem{brin} 
Brin S, Page L
\newblock (1998)
\newblock {The anatomy of a large-scale hypertextual Web search engine}
\newblock Computer Networks and ISDN Systems {\bf 30}, 107 

\bibitem{meyer} 
Langville AM, Meyer CD
\newblock (2006)
\newblock {Google's PageRank and beyond: the science of  search engine rankings}
\newblock Princeton University Press, Princeton 

\bibitem{rmp2015} 
Ermann L, Frahm KM, Shepelyansky DL
\newblock (2015)
\newblock {Google matrix analysis of directed networks}
\newblock Rev. Mod. Phys. {\bf 87}, 1261

\bibitem{wtn1} 
Ermann L, Shepelyansky DL
\newblock (2011)
\newblock {Google matrix of the world trade network}
\newblock Acta Physica Polonica A {\bf 120}, A158 

\bibitem{wtn2} 
Ermann L, Shepelyansky DL
\newblock (2015)
\newblock {Google matrix analysis of the multiproduct world trade network}
\newblock Eur. Phys. J. B {\bf 88}, 84 

\bibitem{wtn3} 
Coquide C, Ermann L, Lages J, Shepelyansky DL
\newblock (2019)
\newblock { Influence of petroleum and gas trade on EU economies 
from the reduced Google matrix analysis of UN COMTRADE data}
\newblock Eur. Phys. J. B {\bf 92}, 171

\bibitem{wtn4} 
Coquide C, Lages J, Shepelyansky DL
\newblock (2020)
\newblock { Crisis contagion in the world trade network}
\newblock Appl. Network Science (Springer) {\bf 5}, 67

\bibitem{serrano07} 
Serrano MA, Boguna M, Vespignani A
\newblock (2007)
\newblock { Patterns of dominant flows in the world trade web}, 
\newblock Journal of Economic Interaction and Coordination {\bf 2(2)}, 111

\bibitem{fagiolo09} 
Fagiolo G, Reyes J, Schiavo S
\newblock (2009)
\newblock { World-trade web: Topological properties, dynamics, and evolution}, 
\newblock Phys. Rev. E {\bf 79}, 036115

\bibitem{he10} 
He J, Deem MW 
\newblock (2010)
\newblock { Structure and response in the world trade network} 
\newblock Phys. Rev. Lett. {\bf 105}, 198701

\bibitem{fagiolo10} 
Fagiolo G, Reyes J, Schiavo S
\newblock (2010) 
\newblock {The evolution of the world trade web: a weighted-network analysis}
\newblock Journal of Evolutionary Economics {\bf 20}, 479 

\bibitem{barigozzi10} 
Barigozzi M, Fagiolo G, Garlaschelli D 
\newblock (2010)
\newblock { Multinetwork of international trade: a commodity-specific analysis}
\newblock Phys. Rev. E {\bf 81}, 046104

\bibitem{debenedictis11} 
De Benedictis L, Tajoli L
\newblock (2011)
\newblock { The world trade network}, 
\newblock The World Economy {\bf 34(8)}, 1417

\bibitem{deguchi14}
\newblock (2014) 
Deguchi T, Takahashi K, Takayasu H, Takayasu M
\newblock { Hubs and authorities in the world trade network using a weighted hits algorithm}, 
\newblock PLoS ONE {\bf 9(7)}, 1 

\bibitem{linux}  
\newblock (2010) 
Chepelianskii AD
\newblock { Towards physical laws for software architecture},
\newblock arXiv:1003.5455 [cs.SE]

\bibitem{wikizzs} 
Zhirov AO, Zhirov OV, Shepelyansky DL
\newblock (2010) 
\newblock { Two-dimensional ranking of Wikipedia articles},
\newblock Eur. Phys. J. B {\bf 77}, 523

\bibitem{greduced} 
Frahm KM, Shepelyansky DL
\newblock (2016) 
\newblock {Reduced Google matrix},
\newblock arXiv:1602.02394[physics.soc]

\bibitem{politwiki} 
Frahm KM, Jaffres-Runser K, Shepelyansky DL,
\newblock (2016) 
\newblock { Wikipedia mining of hidden links between political leaders},
\newblock Eur. Phys. J. B {\bf 89}, 269

\bibitem{wikicountires} 
El Zant S, Jaffres-Runser K, Shepelyansky DL,
\newblock (2018) 
\newblock { Capturing the influence of geopolitical ties from Wikipedia with reduced Google matrix},
\newblock PLoS ONE {\bf 13(8)}, e0201397

\bibitem{wrwu2017} 
Coquide C, Lages J, Shepelyansky DL 
\newblock (2019) 
\newblock { World influence and interactions of universities from Wikipedia networks},
\newblock Eur. Phys. J. B {\bf 92}, 3

\bibitem{wikipharma} 
Rollin G, Lages J, Serebriyskaya TS, Shepelyansky DL
\newblock (2019) 
\newblock { Interactions of pharmaceutical companies with 
world countries, cancers and rare diseases from Wikipedia network analysis},
\newblock PLoS ONE {\bf 14(12)}, e0225500

\bibitem{zinprotein1} 
Lages J, Shepelyansky DL, Zinovyev A
\newblock (2018) 
\newblock { Inferring hidden causal relations between 
pathway members using reduced Google matrix of directed biological networks}
\newblock PLoS ONE {\bf 13(1)}, e0190812

\bibitem{zinprotein2} 
Zinovyev A, Czerwinska U, Cantini L, Barillot E, Frahm KM, Shepelyansky DL
\newblock (2020) 
\newblock {Collective intelligence defines biological 
functions in Wikipedia as communities in the hidden protein connection network},
\newblock PLoS Comput Biol {\bf 16(2)}, e1007652


\bibitem{escaith} 
Kandiah V, Escaith H, Shepelyansky DL
\newblock (2015) 
\newblock {Google matrix of the world network of economic activities}, 
\newblock Eur. Phys. J. B {\bf 88}, 186

\end{thebibliography}
\end{document}